\documentclass{ws-ijmpa}
\usepackage{epsfig}
\usepackage{rotate}
\usepackage{amssymb}
\usepackage{amsmath}

\newcommand{\eg}{{\it{e.g.}}}
\newcommand{\ie}{{\it{i.e.}}}

\newcommand{\ops}{\ensuremath{\mathrm{o\text{-}Ps}}}
\newcommand{\oPs}{\ensuremath{\mathrm{o\text{-}Ps}}}
\newcommand{\pps}{\ensuremath{\mathrm{p\text{-}Ps}}}
\newcommand{\nplus}{\ensuremath{N_+}}
\newcommand{\nminu}{\ensuremath{N_-}}
\newcommand{\kone}{\ensuremath{\hat{k}_1}}
\newcommand{\ktwo}{\ensuremath{\hat{k}_2}}
\newcommand{\kthr}{\ensuremath{\hat{k}_3}}
\newcommand{\shat}{\ensuremath{\hat{S}}}
\newcommand{\Bvec}{\ensuremath{\vec{B}}}
\newcommand{\ccp}{\ensuremath{{C_{CP}}}}
\newcommand{\au}{\ensuremath{A_{u}}}
\newcommand{\ap}{\ensuremath{A_{p}}}

\long\def\symbolfootnote[#1]#2{\begingroup%
\def\thefootnote{\fnsymbol{footnote}}\footnote[#1]{#2}\endgroup} 

\begin{document}

\markboth{M.~Felcini}
{A TEST OF CP SYMMETRY IN POSITRONIUM}

\catchline{}{}{}

\title{A TEST OF CP SYMMETRY IN 
POSITRONIUM
}

\author{\footnotesize Marta FELCINI\footnote{E-mail: marta.felcini@cern.ch; 
Postal address: 
CERN Physics Department, CH-1211 Geneva}}

\address{Institute for Particle Physics, ETH-Zurich\\
CH-8097 Zurich, Switzerland}

\maketitle

\pub{February 5, 2004\footnote{To be published in the Proceedings of the 
``Workshop on Positronium Physics'', ETH-Zurich, May 30-31, 2003. }}{}

\begin{abstract} 
The aim of this CP symmetry test in positronium 
is to measure the CP violation amplitude 
parameter \ccp. This is derived from the measurement of
the asymmetry in an angular distribution of the photons from 
the decay of the ortho-positronium in a magnetic field. 
The Standard Model prediction for \ccp\ is a value of 
the order of $10^{-9}$.
Thus the observation of a larger \ccp\ value  would be 
signal of physics beyond the Standard Model.
A previous measurement has found \ccp\ consistent with zero, with an
uncertainty of $\sim  10^{-2}$. We have investigated 
the possibility of using the existing ETHZ-INRM-IN2P3 BGO crystal
detector, set-up for positronium physics studies, 
to improve the sensitivity on the \ccp\ measurement. 
Preliminary calculations indicate 
that, using such an apparatus, with some modification, in a 
magnetic field of 4 kGauss,  
\ccp\ could be measured with an uncertainty in the range
between   $\sim 10^{-4}$ and $\sim 10^{-3}$, 
depending mainly on the 
uncertainty  in the asymmetry measurement and the angular resolution 
of the photon detectors.
If  \ccp\ is less than $\sim 10^{-4}$, the experimental technique outlined
here appears to be inadequate to observe a CP violating effect and 
new techniques
or different observables must be exploited for better sensitivity.

\keywords{Positronium; CP symmetry test; BGO crystal detector.}
\end{abstract}

\section{Introduction}	
In the context of the Standard Model of particle physics, violation of 
the discrete symmetry CP (C=charge conjugation, P=parity operation) and 
time reversal T arise in the quark sector only through 
the Cabibbo-Kobayashi-Maskawa (CKM) matrix~\cite{kobmas}. 
A number of measurements,
mainly in the K and B meson sectors, allow to constrain the mixing angles and
the phase of this matrix and to check the consistency of the model
(see \eg\ Ref.~\cite{Rosner:2000bh} for a review).
This model, however, does not   
explain the observed pattern of mixing and CP violation 
in the quark sector.
The situation in the lepton sector is different. Within the Standard Model,
the lepton sector shows invariance under the discrete symmetry 
transformations CP and T. In the Standard Model, CP violating effects 
in a lepton system can arise only through higher order corrections from 
the CP violation in the quark sector, and they are expected to be extremely 
small. However, the recent observation
of neutrino oscillations implies the existence of a lepton mixing matrix
the so-called Pontecorvo-Maki-Nagakawa-Sakata (PMNS) 
matrix~\cite{pontecorvo}, 
which is analogous to the CKM matrix for quarks. 
If it contains non-zero CP-violating
phases, then it can induce CP-violating effects in the neutrino and the 
charged lepton sectors. If the PMNS matrix is the only source 
of CP violation in the lepton sector, then neutrino oscillations are likely 
to be the only manifestation of CP violation in the lepton sector, 
as the CP-violating effects induced on charged leptons by the PMNS matrix
are expected to be unobservably small (see \eg\ Ref.\cite{lavignac} 
for a recent review) .  
Thus, experimental evidence for genuine CP violation and/or T violation in a
charged lepton system would be signature of new interactions among leptons, 
with an associated leptonic mixing matrix containing CP violating phases.  
A compilation of present experimental results of CP and T symmetry tests for 
leptons can be found in Ref.~\cite{pdg}.

It was realized since several years  that the positronium, 
an electron-positron bound state, can serve as a testing 
ground for different
discrete symmetries as C, P, T, CP and CPT in the charged 
lepton sector (see Ref.~\cite{krasnikov} for a theoretical review).
Previous and new experimental results on C and CPT symmetry tests
in positronium have been reported~\cite{vetter}. 

\begin{figure}[tb]
\begin{center}
\mbox{\epsfig{figure=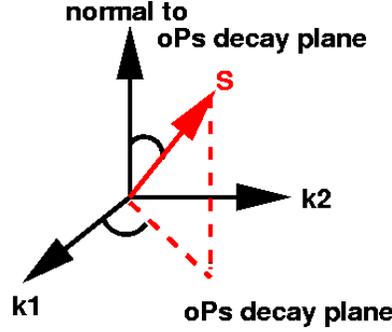,width=0.4\textwidth}}
\caption{Definition of the geometry and of the quantities 
used for the experimental 
tests of CP symmetry in the positronium system. 
The vectors \kone\ and \ktwo,
the directions of the first and second highest energy photons, 
respectively, identify 
the decay plane of the \ops. The vector S
indicates the \ops\ spin. }
\label{geometry}
\end{center}
\end{figure}

\begin{figure}[tb]
\begin{center}
\mbox{\epsfig{figure=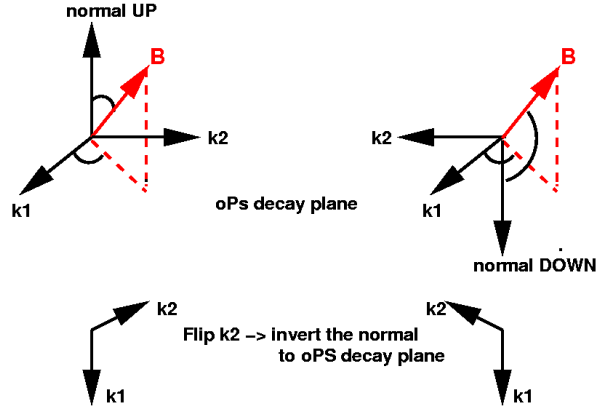,width=0.4\textheight}}
\caption{Definition of \nplus\ and \nminu\ events: \nplus\ is the number 
of events
where the normal to the \ops\ decay plane forming an angle smaller than 
90 degrees with the B field vector (defined us normal UP in the picture). 
Flipping \ktwo, with respect to \kone, 
as shown in the picture, inverts the normal to the \ops\ decay plane
(defined as normal DOWN in the picture).}
\label{definition}
\end{center}
\end{figure}

In this paper we concentrate on CP symmetry tests 
in the positronium system. A previous experiment is described
in Ref.~\cite{Skalsey:vt}. The experiment aims to measure the angular
correlation between the positronium  spin and the momenta of the 
photons from the positronium decay.
As proposed in Ref.~\cite{Bernreuther:tt}, 
if CP violating interactions are acting in the positronium system, 
then (at least) one CP violating term must contribute to the positronium 
lagrangian. This term should be proportional to:
\begin{equation}
(\shat\cdot\kone)(\shat\cdot\kone\times\ktwo)\ .
\label{equone}
\end{equation}
Here $\shat$ is the positronium spin, while \kone\ and \ktwo\ are 
the unit vectors in the directions 
of the highest and second highest energy photons from the three-photon decay 
of positronium. Indeed, the quantity in eq.~\ref{equone} 
is non-zero only for the spin 1 state of the positronium, 
the ortho-positronium (\ops), which decays into three photons.  
The \kone\ and \ktwo\ unit vectors identify a plane (see
Fig.~\ref{geometry}). The first product in eq.\ref{equone} 
is the cosine of the 
angle between the \ops\ spin and the highest energy photon, which we
define  as $\cos{\theta_{1}}$. The second product 
in eq.~\ref{equone} is the cosine of the angle between the 
\ops\ spin and the normal to the \kone-\ktwo\ plane 
(see Fig.~\ref{geometry}),
which we define as $\cos{\theta_{2}}$.
Then the quantity in eq.~\ref{equone}, can be rewritten as
$ \cos{\theta}\equiv\cos{\theta_{1}}\cos{\theta_{2}}$.
If there is a CP violating term in the lagrangian, then the measured 
number of events as a function of $\cos{\theta}$ should be described by
\begin{equation}
\ensuremath{N(\cos{\theta})=N_0(1+\ccp\cos{\theta})\ ,}
\end{equation}
with the CP violation amplitude parameter, \ccp, different from zero.   
Within the Standard Model, \ccp\ in the positronium system is expected to be
very small, of the order of $10^{-9}$ (see Ref.~\cite{Skalsey:kw}).
The measured distribution  $N(\cos{\theta})$ 
should show an 
asymmetry in $\cos{\theta}$ so that 
$\ensuremath{N(\cos{\theta_+})-N(\cos{\theta_-})}=2N_0\ccp\cos{\theta}$ 
for $\cos{\theta_+}=-\cos{\theta_-}=\cos{\theta}$. 
The quantity \ccp\ can be determined by measuring   
the rate of events \nplus\
for a given $\cos{\theta_+}=\cos{\theta}$ and \nminu\ for 
$\cos{\theta_-}=-\cos{\theta}$. In practice, \nplus\ is the 
number of events in which \ktwo\ forms an angle $\theta_{12}$,
smaller than 180 degrees, with \kone\ and 
the B field forms an angle $\theta_{2}$ 
smaller than 90 degrees with the normal to the \ops\ decay plane 
(see Fig.~\ref{definition}).
In the \nminu\ events, \ktwo\ forms an angle $2\pi-\theta_{12}$
with \kone\ and the B field forms an angle $2\pi-\theta_{2}$ 
with the normal to the \ops\ decay plane. 
For these events the normal to the \ops\ decay is reversed with respect
to the \nplus\ events, by flipping the direction of \ktwo\ specularly 
with respect to \kone (see Fig.~\ref{definition}).
Then the asymmetry
\begin{equation}
A=\frac{(\nplus-\nminu)}{(\nplus+\nminu)}=\ccp\cos{\theta}
\label{equasy}
\end{equation}
allows to derive the experimental value of \ccp.

\section{Experimental method}
The measurement of the asymmetry $A$ implies that 
$\cos{\theta}$ in eq.~\ref{equasy} is a well defined quantity 
in the experiment.
In turn, this implies that  the spin $S$ direction is defined.
This direction can be selected using an external magnetic field.

\begin{figure}[tb]
\begin{center}
\mbox{\epsfig{figure=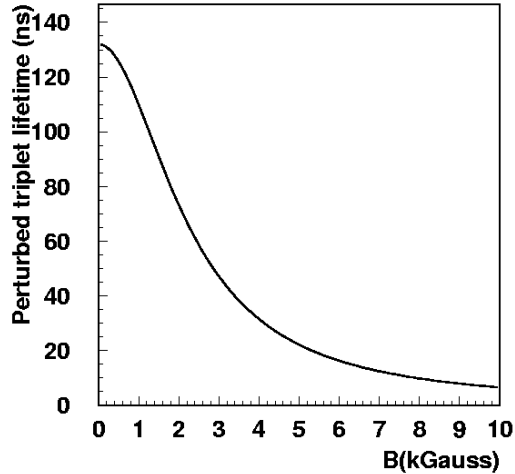,width=0.6\textwidth}}
\caption{The lifetime of the $m=0$ Ps states perturbed by an external 
magnetic field, as a function of the magnetic field intensity B. }
\label{perttrip}
\end{center}
\end{figure}

Let us recall the Ps spin states: 
\begin{itemize}
\item the singlet, with S=0, m=0, for the parapositronium (\pps);
\item the triplet, with S=1, m=$1,0,-1$, for the orthopositronium (\ops).     
\end{itemize}
An external magnetic field \Bvec\ can be used to align the \ops\ spin
parallel (m=1), perpendicular (m=0) or antiparallel (m=$-1$)
to the field direction.
However, the magnetic field does not only align the spin, 
but also perturbs and 
mixes the m=0 states. Thus, two new states are possible for the Ps system:
the perturbed singlet and the perturbed triplet m=0 states. 
Their lifetimes depend on the \Bvec\ field intensity.
The perturbed singlet state has a lifetime shorter than 1 ns (as the 
unperturbed one), which is not relevant in the measurement described here,
because too short compared to the typical detector time resolution of 1 ns. 
The perturbed triplet lifetime
as a function of the magnetic field intensity B is shown in
Fig.~\ref{perttrip} (calculated using the lifetime formula 
in Ref.~\cite{Skalsey:kw}). 
One sees that for values of B of few kGauss the lifetime of the triplet m=0 
state can be substantially reduced with respect to the unperturbed lifetime. 
Thanks to this effect, applying an external magnetic field, 
it is possible to separate the m=0 from the m=$\pm 1$ states, by the 
different lifetimes of the perturbed m=0 and unperturbed m=$\pm 1$ states.
The values of the B field can be optimized for maximum separation: it is found
that for B=4 kGauss, corresponding to a perturbed lifetime of 30 ns, the
separation is optimal. 

\begin{figure}[tb]
\begin{center}
\mbox{\epsfig{figure=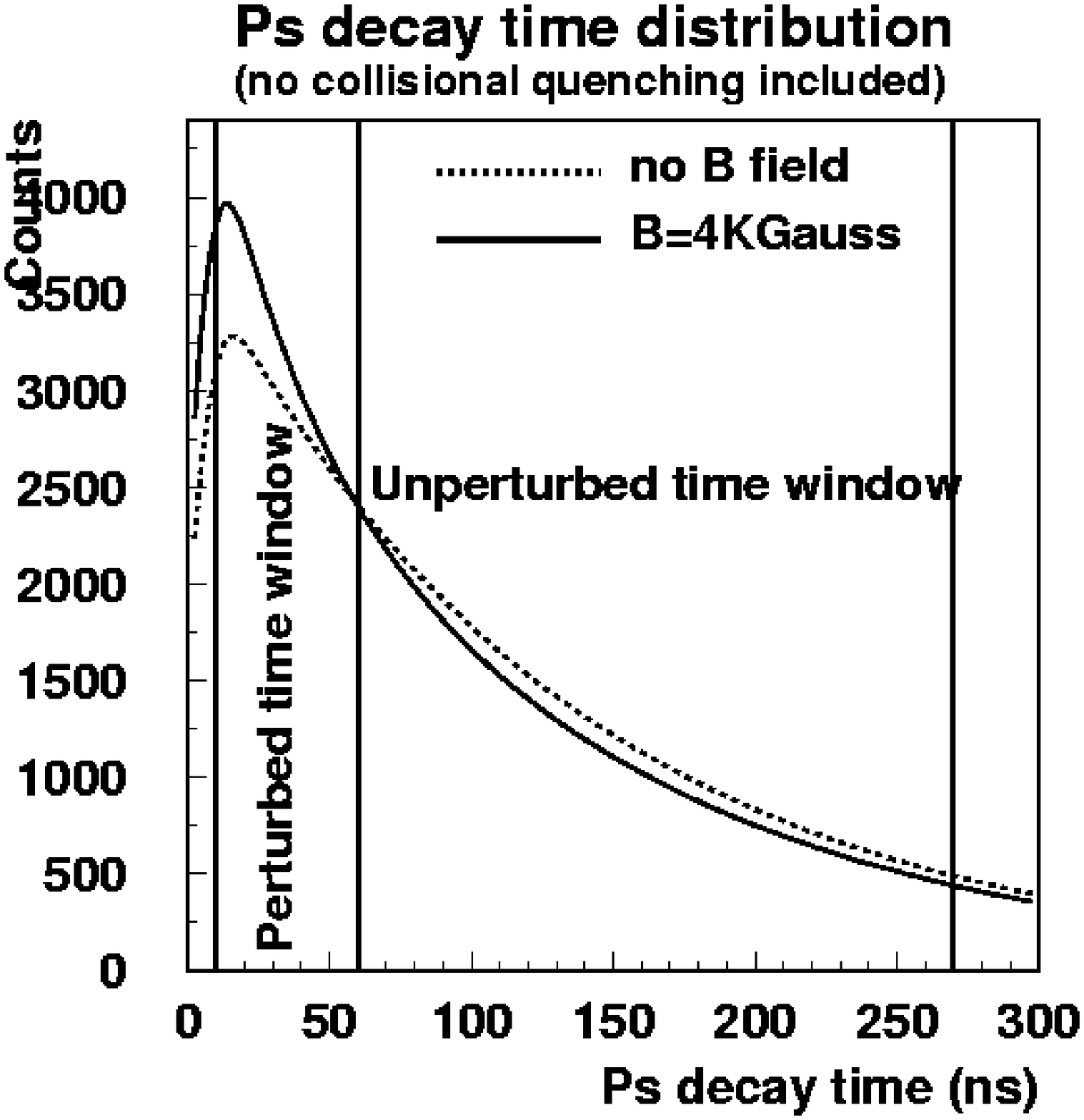,width=0.6\textwidth}}
\caption{The positronium decay time spectrum without and with an 
external magnetic field of 4 kGauss, corresponding to a perturbed 
lifetime of the triplet $m=0$ state of 30 ns 
(see Fig.~\ref{geometry}). For this calculation, the population of each
of the four states, one singlet (m=0) and three triplet (m=1,0,-1) 
states, is taken to be the same. The m=0 states are mixed by the 
magnetic field and their lifetime modified, depending on the 
magnetic field intensity.
The unperturbed singlet lifetime is $\tau_s=0.125$ ns, while the perturbed 
singlet lifetime for B=4 kGauss is 0.522 ns. This value is used for
the lifetime of the singlet m=0 state, in the presence of the magnetic field. 
The unperturbed triplet lifetime is
$\tau_\ops=132$ ns, while the perturbed triplet m=0 lifetime 
for B=4 kGauss is 30 ns. Note that the triplet m=$\pm 1$ states 
are unperturbed thus they have a lifetime of 132 ns, without or with the
magnetic field.
}
\label{decaytime}
\end{center}
\end{figure}

Fig.~\ref{decaytime} shows the positronium decay time spectrum without 
and with an external magnetic field of 4 kGauss, corresponding to a perturbed 
lifetime of the triplet $m=0$ state of 30 ns 
(see Fig.~\ref{perttrip}). For this calculation, the population of each
of the four states, one singlet (m=0) and three triplet (m=$1,0,-1$) 
states, is taken to be the same. The m=0 states are mixed by the 
magnetic field and their lifetime modified, depending on the 
magnetic field intensity.
The unperturbed singlet lifetime is 0.125 ns, while the perturbed 
singlet lifetime for B=4 kGauss is 0.522 ns. This value is used for
the lifetime of the singlet m=0 state, in the presence of the magnetic field. 
The unperturbed triplet lifetime used in the calculation is 
132 ns, as measured in SiO$_2$ aerogel target of 
Ref.~\cite{Badertscher:2002nh}.
The perturbed triplet m=0 lifetime 
for B=4 kGauss is 30 ns. Note that the triplet m=$\pm 1$ states 
are unperturbed, thus they have a lifetime of 132 ns, without or with the
external magnetic field.

The measurement of the asymmetry $A$ is actually done in the following way.
The direction and intensity of the B field are fixed. The \kone\ and
\ktwo\ detectors are also fixed. In this way $\cos{\theta}$ is defined.
For each event, the Ps decay time and the energies of the three photons 
from the \oPs\ decay are measured.
The off-line analysis requires that the highest energy photon be in 
the \kone\ detector within an energy range $\Delta E_1=E_1^{max}-E_1^{min}$. 
The second highest energy
photon must be recorded in the \ktwo\ detector within an 
energy range $\Delta E_2=E_2^{max}-E_2^{min}$.   
Then the \nplus\ and \nminu\ events are counted to determine 
the asymmetry in eq.~\ref{equasy}. 

The measurement of the asymmetry $A$ is performed for both the 
perturbed m=0 states, selecting events with Ps decay time between 10 and 60 ns
(``perturbed time window'' in Fig.~\ref{decaytime}),
and   for the unperturbed m=$\pm 1$ states, selecting 
events with Ps decay time between 60 and 270 ns 
(``unperturbed time window'' in Fig.~\ref{decaytime}).
Let us define \au\ and \ap\ the asymmetry measurements for the perturbed 
and unperturbed states, respectively.  
In this way two independent asymmetry measurements are available.
Furthermore it is easy to demonstrate~\cite{Skalsey:vt} that 
if $\au=\ccp\cos{\theta}$ 
then $\ap=-\ccp\cos{\theta}$.
Thus, the quantity 
\begin{equation}
A=(\au-\ap)/2
\label{equasyfin}
\end{equation} 
provides a measurement of \ccp\ which is free of decay-time-independent 
systematics.
Decay-time-dependent systematics, on the contrary, are not 
canceled out in eq.~\ref{equasyfin} and they must be considered and 
included in the total uncertainty on $A$, 
as explained in the next section. 

The measured asymmetry is related to the \ccp\ parameter by the simple
formula
\begin{equation}
A=\ccp Q\ .
\label{equsim}
\end{equation}
The quantity Q is ideally the $\cos{\theta}$ 
value in eq.~\ref{equasy}.
However, due to the experimental uncertainty on the 
$\cos{\theta}$ value, to the backgrounds affecting the asymmetry measurement,
and to other uncertainties, the measured asymmetry value 
is actually reduced with 
respect to an ideal case with no uncertainties. Comparing eq.~\ref{equsim}
with eq.~\ref{equasy}, 
the quantity $Q$ can  be written as 
\begin{equation}
Q=f\cos{\theta}\ ,
\end{equation}
with $f<1$ in a real experiment. 
The quantity $Q$ is the product of several factors, 
accounting for effects and uncertainties deteriorating the asymmetry 
measurement with respect to the case of an ideal detectors.  
It can be determined, for a given detector geometry, by 
a Monte Carlo simulation and by measurements. 
The uncertainty on \ccp\ is determined by the uncertainties on the measured $A$
value and on the evaluated Q value. The uncertainty on \ccp\ determines the 
level of sensitivity of the experiment to a hypothetical CP violating 
interaction.

\section{A new CP symmetry test experiment}

\begin{figure}[tb]
\begin{center}
\mbox{\epsfig{figure=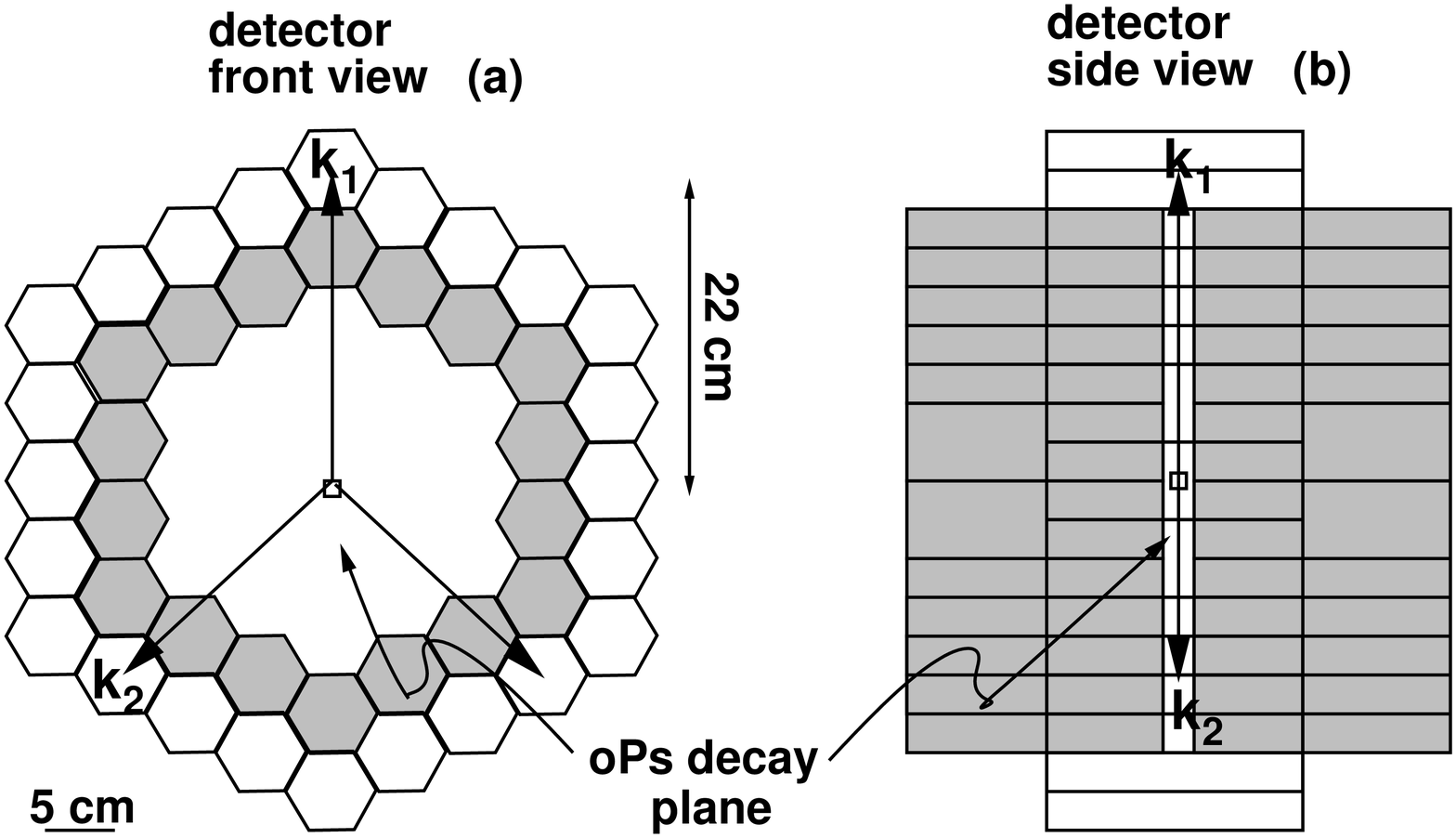,width=0.95\textwidth}}
\caption{Schematic view of the  BGO crystal barrel calorimeter
used to detected the photons from
the \ops\ decay: (a) detector front view  and definition of the 
\kone\ and \ktwo\ vectors, for this arrangement: the outer crystal ring
is used for the photon measurements, the inner crystal rings (shaded region) 
are used as veto; 
(b) detector side view, showing the crystal outer ring for photon detection
and the inner rings (shaded region) for veto:
only the photons detected by the outer crystals
in the window between the veto crystals are considered for 
the \ccp\ measurement.
This arrangement allows to improve the crystal angular resolution for 
photon detection, as compared to the case where no veto is used.
}
\label{newdetec}
\end{center}
\end{figure}

\begin{figure}[htb]
\begin{center}
\mbox{\epsfig{figure=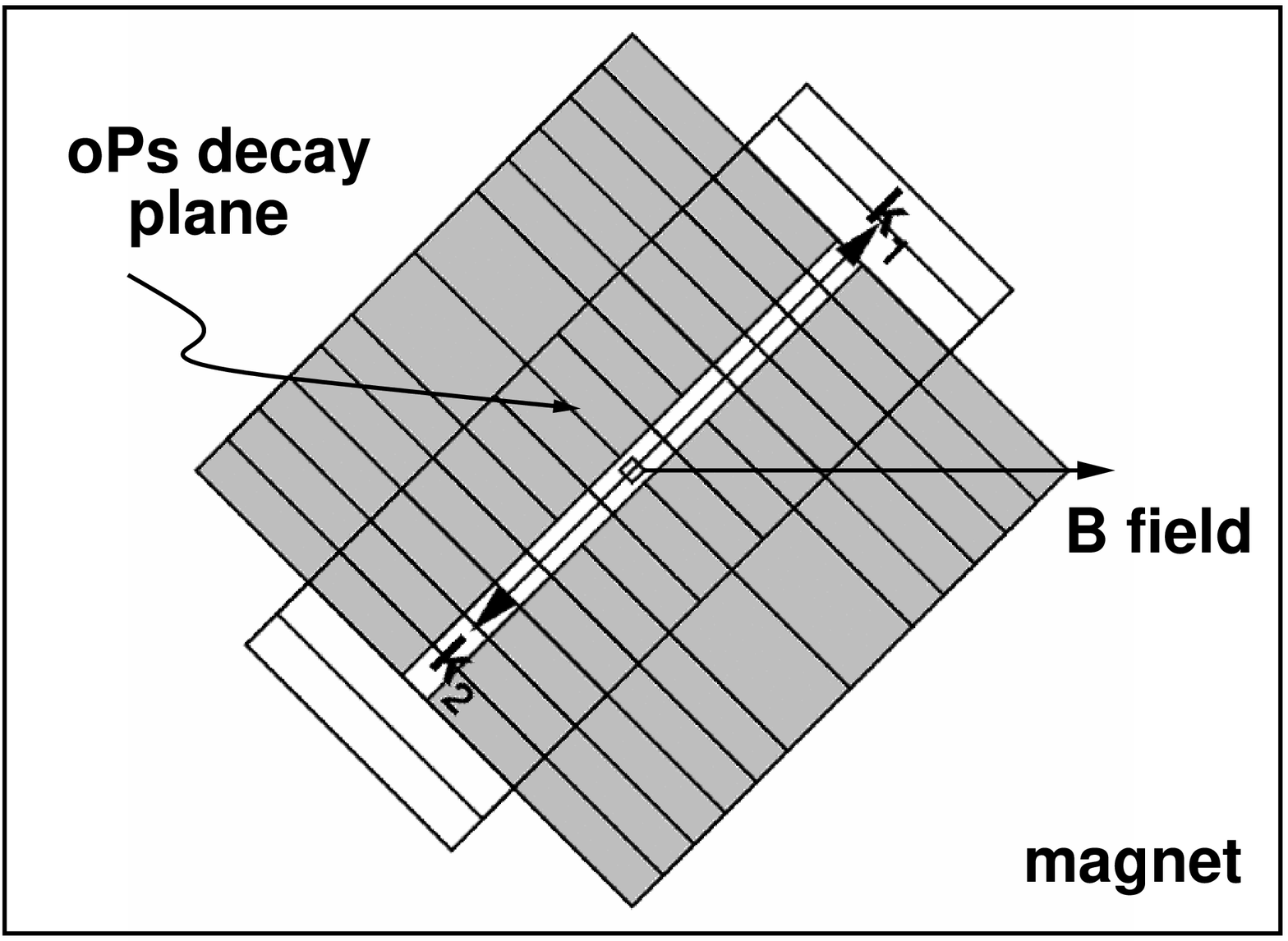,width=0.75\textwidth}}
\caption{Schematic side view of the crystal barrel placed in a solenoidal
magnet, generating a magnetic field in the direction shown in the picture.
The direction of the B field with respect to \kone\ and to the 
the normal to the \kone-\ktwo\ plane is chosen to be 
45 degrees, to maximize the sensitivity of the \ccp\ measurement.}
\label{detectorinmagnet}
\end{center}
\end{figure}

In the previous experiment to test CP invariance~\cite{Skalsey:vt},
the \ccp\ value is measured to be
\begin{equation}
\ccp = -0.0056\pm 0.0154\ .
\label{equmea}
\end{equation}
The total uncertainty includes both statistical and systematic uncertainties.  
This result is obtained by the asymmetry measurement and the evaluated value 
of $Q$ (see eq.~\ref{equsim}). 
The evaluated Q value (called analyzing power $S_{an}$ in~\cite{Skalsey:vt}) 
is $Q=0.072\pm 0.15$. The value of $\cos{\theta}$, for $\theta_1=55^o$ and
$\theta_2=45^o$, is $\cos{\theta}=0.41$, resulting into 
$f=Q/\cos{\theta}=0.17$, about a factor of 6 smaller than for a 
perfect detector ($f=1$). It implies that in a new  
experiment, at the best $Q$ can be increased by a factor of $\sim 6$, 
as compared to the previous experiment.
The quantity Q is evaluated as the product of several factors: 
the dominant factors are 
related to the (decay-time dependent) background and 
to the uncertainties on $\theta_1$ and $\theta_2$
values, determined by the photon detector angular 
resolution~\cite{Skalsey:vt}. 

The final asymmetry measurement gives:
\begin{eqnarray}
A_{final}=-0.0004\pm 0.0010(stat.)+\pm 0.0004(syst.1)\pm 0.0001(syst.2)
=\nonumber \\
-0.0004\pm 0.0011\ .
\label{oldaresult}
\end{eqnarray} 

The statistical uncertainty on $A_{final}$, with few  $10^6$ 
events collected, amounts to less
than $10^{-3}$. The systematic uncertainty on $A_{final}$ 
is determined by two effects, which are both decay-time-dependent, \ie\
affecting differently the measurements made in the unperturbed and
perturbed time windows (recall definition of $A$  by eq.~\ref{equasyfin}).

The first systematic uncertainty on $A_{final}$ 
in eq.~\ref{oldaresult} is due to 
the subtraction of background induced mainly by events
with two back-to-back annihilation photons (from \eg\ collisional 
quenching of unperturbed triplet Ps or magnetic quenching of 
perturbed triplet Ps), mimicking a $3\gamma$ event.
This is the case if one of the 511 keV
photon from the two-photon decay
makes Compton scattering and deposits in the \ktwo\ detector an 
energy in the required $\Delta E_2$ range. 
The second systematic uncertainty is introduced by the correction 
for diffusion of the Ps atoms during their lifetime in the positronium
formation region~\cite{Skalsey:vt}.

An additional systematic uncertainty  
on the $A_{final}$ value (not included in eq.~\ref{oldaresult})
may be introduced by the effect of 
the shadowing of the crystals by the coils of the permanent magnet 
(see Fig.~1 in Ref.~\cite{Skalsey:vt}), used to create the magnetic field
over the volume of the \ops\ forming region.
To obtain a magnetic field of 4 to 5 kGauss, as needed, there are two 
options: either the 
use of a small permanent magnet, generating a field just 
around the Ps forming region, or a large magnet, capable to contain the
photon detector. The first option 
has the advantage that the magnetic field affects practically only the Ps
forming region. The disadvantage 
is the effect of the shadowing of the crystals
by the magnet coils.
The second option (large magnet) is envisaged for a new CP 
symmetry test experiment outlined here. 
The advantage of this option is that the detector is inside the 
magnetic field, so there is no shadowing of the crystals by the magnet coils.
In addition the field uniformity is expected to be better than $\pm 1\%$.
The disadvantage of this option is that conventional photomultipliers cannot 
be used inside the strong magnetic field. It is instead possible to use
\eg\ large area avalanche photodiodes (see \eg\ Ref.~\cite{Renker:st} and 
references therein). 
   
The new detector set-up is sketched in Fig.~\ref{newdetec}.
The photon detector is made of BGO crystals. They have an hexagonal cross 
section with an inner diameter of 5.5 cm and a length of 20 cm. 
The crystals are the same  
as those used in the ETHZ-INRM-IN2P3 experiment of 
Ref.~\cite{Badertscher:2002nh} 
and expected to be used in the experiments proposed in  Ref.~\cite{gninenko}
and  Ref.~\cite{crivelli}. About hundred of these crystals are available,
from a previous experiment(see Ref.~\cite{Barnett:ew}).
The positronium formation and tagging methods
can be similar to those described in Ref.~\cite{Badertscher:2002nh}.
The crystals surround the positronium formation and decay region.
In the set-up shown in Fig.~\ref{newdetec}, 24 crystals are 
used for a barrel detector. 
The advantage of having a complete crystal barrel 
surrounding the Ps decay region is that several asymmetry measurements can 
be done for different photon angles, without any intervention on the detector
between the measurements. The different asymmetry measurements,
when combined together,  
are expected to result in a reduction of systematic uncertainties,
due to geometrical effects, inducing fake asymmetries. 

The crystals are placed $\sim 20$ cm from the \ops\ decay 
region\footnote{The overall size of the detector is determined 
by the number of available crystals and the space available 
for the detector inside the magnet.}. 
Thus the solid angle covered by one crystal is much larger than 
the $\sim$2\% of $4\pi$
used for the measurement in~\cite{Skalsey:vt}
(indeed the photon detectors where mounted in
lead shields to reduce the photon detection area). 
Such a poor angular resolution for the present photon detectors
would result into an unacceptably large uncertainty on the photon angle
measurements. Thus, for a comparable or better sensitivity, than in the
previous experiment, we need at least a comparable photon angular resolution.
A way to improve this resolution is by reducing the photon measuring
area. This could be realized by reducing the crystal size, 
cutting the crystals into smaller ones.
Allthough technically feasible, this is not envisageable in the short term,
because the crystals are being used also for other experiments, where the
bigger size is preferred. An alternative way to reduce the 
photon measuring area is by shielding part of the 
(rectangular) face of the crystal exposed to the photons, so that 
only a small window is actually used to detect the photons for the
\ccp\ measurement. 
An efficient shield can be realized by additional BGO crystals, in
a set-up as the one shown in  Fig.~\ref{newdetec}. Two inner rings, each
of 18 BGO crystals (shaded area), can be used as a veto. A small distance
$d$ is left among the two inner rings (Fig.~\ref{newdetec}(b)). 
In this way, if $l$ is the shorter 
side of the crystal rectangular face exposed to photons, a 
photon measuring window of area $l\times d$ is defined on this  
face of the outer crystal.  
If a signal is detected in the veto crystals, then the event is rejected.
The event is accepted by the off-line analysis if three photons, \kone,
\ktwo\ and \kthr\ are recorded in the  photon measuring window of the 
respective outer ring crystals, with energies $E_1$, $E_2$ and $E_3$
in the required energy ranges.   
An angular resolution comparable to the previous experiment is obtained if 
the distance between the inner veto rings is about 1 cm.
Then the photon measuring window on the outer crystals has an area
of $l\times d=5.5\times 1$ cm$^2$, resulting into
$(5.5/20^2)\sim 1.4\%$ of 4$\pi$. 

The detector is placed inside a solenoidal magnet 
(see Fig.\ref{detectorinmagnet}), generating a uniform and constant magnetic
field of 4 kGauss, along the direction of the magnet longitudinal axis. 
The crystal calorimeter must be held in 
place by a rigid mechanical structure. 
The B field direction must not be perpendicular to \kone\ and not lying 
in the  \kone-\ktwo\ plane, for a non-zero value of the CP violating
term (eq.\ref{equone}). The $\cos{\theta}$ value is maximum (best 
sensitivity to \ccp) if the 
B vector is lying in the plane identified by \kone\ and the normal to 
the \kone-\ktwo\ plane, and forming an angle of $45^o$ with \kone.
In this case $\cos{\theta}=0.5$. 
  
\section{Estimate of the detector sensitivity}
To understand how to optimize the sensitivity of the detector we must consider
the uncertainties affecting the \ccp\ measurement. They are the uncertainties
on the 
$A$ asymmetry measurement and on $Q$. In the following, we quantify the
uncertainty on \ccp\ as a function of the quantities $A$ and $Q$ and their 
uncertainties.
Let us define $\Delta\ccp$, $\Delta A$ and $\Delta Q$ the uncertainties 
 on \ccp, $A$ and $Q$, respectively. 
Then, the relative error on \ccp\ is given by:
$$ | \frac{\Delta\ccp}{\ccp}|^2=|\frac{\Delta A}{A}|^2+
|\frac{\Delta Q}{Q}|^2\ .$$  
which can be written as:
$$| \frac{\Delta\ccp}{\ccp}|^2=\frac{1}{Q^2}\Big[ |\frac{\Delta A}{\ccp}|^2
+|\Delta Q|^2 \Big]\ . $$
The uncertainty on $\ccp$ is then:
$$\Delta\ccp=\frac{1}{Q}\Big [ |\Delta A|^2 +\ccp^2|\Delta Q|^2 \Big ]^{1/2}$$
which for $\ccp<<1$ can be approximated by
\begin{equation}
\Delta\ccp\approx \frac{|\Delta A|}{Q}\ ,
\label{deltac}
\end{equation}
where $\Delta A$ is the total uncertainty, combining both the 
statistical and systematic uncertainties.

The statistical uncertainty on $A$ is determined by the event statistics: 
$$\Delta A_{stat} \sim \sqrt{2/(\nplus+\nminu)}\ .$$ 
For $\nplus+\nminu$ of the order 
of $10^{10}$, then the statistical uncertainty is 
$\Delta A_{stat}\sim 10^{-5}$.
A systematic uncertainty is resulting from the subtraction of the 
two-photon annihilation background which, as explained in the previous Section,
affects differently the asymmetry measurements 
in the perturbed and unperturbed time windows. 
To see how a decay-time-dependent background affects the asymmetry 
measurement, let us write the asymmetry expression including a 
decay-time-dependent background component. 
This is given by
\begin{equation}
A^{meas}=\frac{\nplus-\nminu+B}{\nplus+\nminu+B}\approx 
A +\frac{B}{\nplus+\nminu}\ .
\end{equation}   
In this equation, $A$ is defined as in eq.~\ref{equasy} and $B$ is the
background counts (the approximate expression is valid for 
$B/(\nplus+\nminu)<<1$).
Combining the asymmetry measurements in the perturbed and 
unperturbed time windows we obtain:
\begin{equation}
\frac{A^{meas}_u-A^{meas}_p}{2}=A+
\frac{B_u}{(\nplus+\nminu)_u}-\frac{B_p}{(\nplus+\nminu)_p}\ .
\label{difference}
\end{equation} 
The background contributions from $B_u$ and $B_p$ can be evaluated
by a Monte Carlo simulation and subtracted to the measured asymmetry. 
The uncertainty on the background evaluation   
($syst.1$ in eq.~\ref{oldaresult}) contributes 
to the total asymmetry uncertainty. 

We find that, for the detector set-up considered here, 
requiring three ``good'' photons
(detected in the windows of the \kone, \ktwo\ and \kthr\
outer crystals),
with $450< E_1 < 550$ keV,  
with $300< E_2 < 400$ keV,  
and 
with $150< E_3 <250$ keV, the probability of 
observing a background event, in the perturbed
time window, where higher background is expected from the two-photon 
annihilation,  is ${B_p}/{(\nplus+\nminu)_p}<10^{-6}$. 
In the unperturbed window we set this background
contribution conservatively to 
zero so that the uncertainty in the difference in eq.~\ref{difference}
is maximized. Thus, we assign a background subtraction systematic
uncertainty of $10^{-6}$, while it was measured to be 
$4\times 10^{-4}$ in the previous experiment.
The requirement of measuring a third photon in the specific energy 
range and angle (while only two photons where measured in~\cite{Skalsey:vt})
results into a reduction of the expected background from 
two-photon annihilation.

The additional systematic uncertainty ($syst.2$ in 
eq.~\ref{oldaresult}), introduced by the correction 
for diffusion of the Ps atoms during their lifetime in the positronium
formation region~\cite{Skalsey:vt}, appears to be the most difficult to 
evaluate and a detailed study is needed to understand how 
to control it at a level lower than the measured $10^{-4}$. 

For what concerns the $Q$ value, entering in eq.~\ref{deltac},
we have seen that its maximum expected value for an ideal detector, 
is a factor of about 6 greater than the measured value of 0.072 in the 
previous detector.
A realistic value of $Q$ for the present detector is obtained 
by considering $\cos{\theta}=0.5$ (was 0.42 in the previous experiment)
and $f\approx 0.3$, improved by a a factor of about two with respect
to the $f$ value in the previous measurement ($f=0.17$). This improvement is
due to the background reduction in this set-up, already considered in the 
discussion of the asymmetry systematic uncertainty from two-photon background
subtraction. 

In summary, with $\Delta A_{stat}\sim 10^{-5}$, 
$\Delta A_{syst1}\approx 10^{-6}$
and in the optimistic scenario where $\Delta A_{syst2}$ can be reduced at 
the $10^{-5}$ level (in comparison to $10^{-4}$ in the previous experiment),
using $Q=0.3\times0.5=0.15$, we get a total error  
$\Delta\ccp\sim 10^{-4}$ (in this case statistical and systematic
 uncertainty are of the same order). 
In a more pessimistic scenario, where $\Delta A_{syst2}$
is not reduced significantly with respect to $10^{-4}$, 
then   $\Delta A_{syst2}$ dominates the total uncertainty,
resulting into $\Delta\ccp\sim 7\times 10^{-4}$.  
  
\section{Conclusions}
The measurement of CP violating effects in the lepton sector would be 
signature of physics beyond the Standard Model. The positronium 
offers the possibility to test the CP symmetry in a charged lepton 
system. The Standard Model predicts for the CP violation amplitude parameter
in positronium, \ccp, a value of the order of $10^{-9}$.  
The present measurement of 
\ccp\ is consistent with zero at the 1\% level of precision. 
We have outlined an experimental set-up to possibly improve on this 
precision. The experiment is based on the use of the 
ETHZ-INRM-IN2P3 BGO crystal calorimeter in a magnetic field of 4 kGauss. 
We have made a preliminary evaluation of the sensitivity to \ccp\ 
expected for such an experiment. 
We find that the sensitivity  
range $\Delta\ccp\sim 10^{-4}-10^{-3}$ can be accessible to this experiment.
The upper end of this range is determined by the systematic 
uncertainty on the asymmetry measurement, dominated by the effect  
of the Ps atom diffusion during its lifetime in the Ps 
formation region. In general, a reduction of  $\Delta\ccp$ 
can be achieved 
by reducing the total uncertainty on the asymmetry measurement and increasing 
the resolving power $Q$. The latter could be obtained by using
a photon detector with a finer angular resolution than for the detector 
presented here.  
However, if \ccp\ is non zero but less than $\sim 10^{-4}$, 
the experimental technique outlined
here appears to be inadequate to observe a CP violating effect and 
new techniques
or different observables must be exploited for better sensitivity.

\section*{Acknowledgments}
I wish to thank P.~Crivelli, S.~Gninenko, T.~Otto and A.~Rubbia for 
useful comments and constructive criticism.

\end{document}